\documentclass[manuscript]{aastex}

\shorttitle{
Simplified solution to orbit determination} 
\shortauthors{
Asada et al.}

\begin{document}


\title{
Simplified solution to determination of a binary orbit} 



\author{H. Asada, 
T. Akasaka 
and 
K. Kudoh
\affil{Faculty of Science and Technology, 
Hirosaki University, \\
Hirosaki 036-8561, Japan}
}







\begin{abstract}
We present a simplified solution to orbit determination 
of a binary system from astrometric observations. 
An exact solution was found by Asada, Akasaka and Kasai 
by assuming no observational errors. 
We extend the solution considering 
observational data. 
The generalized solution is expressed in terms of elementary
functions, and therefore requires neither iterative nor numerical methods. 
\end{abstract}





\section{Introduction}

In the seventeenth century, Kepler discovered the laws of 
the motion of celestial objects. 
Since then, a fundamental problem has been how to determine 
the orbital elements (and the mass) of a binary from observational 
data of its positions projected onto a celestial sphere. 
In fact, astrometric observations play an important role in astronomy 
through determining a mass of various unseen celestial objects 
currently such as a massive black hole (Miyoshi et al. 1995),  
an extra-solar planet (Benedict et al. 2002) 
and two new satellites of Pluto (Weaver et al. 2006). 

The orbit determination of {\it resolved double stars} was solved 
first by Savary in 1827, secondly by Encke 1832, thirdly by 
Herschel 1833 and by many authors including Kowalsky, Thiele and Innes 
(Aitken 1964 for a review on earlier works; for the
state-of-the-art techniques, e.g, Eichhorn and Xu 1990, 
Catovic and Olevic 1992, Olevic and Cvetkovic 2004). 
Here, resolved double stars are a system of two stars both 
of which can be seen. 
The relative vector from the primary star to the secondary is 
in an elliptic motion with a focus at the primary. 
This relative vector is observable because the two stars are seen. 
On the other hand, an {\it astrometric binary} is a system of 
two objects where one object can be seen but the other cannot 
like a black hole or a very dim star. In this case, it is impossible 
to directly measure the relative vector connecting the two objects, 
because one end of the separation of the binary, namely 
the secondary, cannot be seen. The measures are made in the position 
of the primary with respect to unrelated reference objects 
(e.g. quasars )whose proper motion is either negligible or known. 

There are two major problems in determination of a binary orbit: 
(1) observational errors and 
(2) the conditional equations which connect the observable 
quantities with the orbital parameters. 
The conditional equations are not only non-linearly coupled 
but also transcendental because of Kepler's equation 
(e.g., Goldstein 1980, Danby 1988, Roy 1988, 
Murray and Dermott 1999, Beutler 2004). 

As a method to determine the orbital elements of a binary, 
an analytic solution in an explicit form has been found 
by Asada, Akasaka and Kasai (2004, henceforth AAK) 
by assuming no observational errors. 
This solution is given in a closed form by requiring 
neither iterative nor numerical methods. 
The purpose of this brief article is to extend the AAK solution 
considering observational data. 
We will summarize the AAK solution in $\S$ 2, 
next extend the solution considering observational data in $\S$ 3, 
and make a computational test of the extended solution 
in $\S$ 4. 

\section{AAK solution} 
Here we briefly summarize our notation and formulation 
for ideal cases without any observational error. 
We consider only the Keplerian motion of a star around 
the common center of mass of a binary system by neglecting 
motions of the observer and the common center in our galaxy. 
Let $(\bar x, \bar y)$ denote the Cartesian coordinates 
on a celestial sphere. 
A general form of an ellipse on the celestial sphere is 
\begin{equation}
\alpha \bar x^2 + \beta \bar y^2 + 2\gamma \bar x \bar y 
+ 2\delta \bar x + 2\varepsilon \bar y = 1 , 
\label{ellipse1}
\end{equation}
which is specified by five parameters 
$(\alpha, \beta, \gamma, \delta, \varepsilon)$ 
since the center, the major/minor axes and the orientation 
of the ellipse are arbitrary. 
Here, we should note that the origin of the coordinates 
is arbitrary in this paper, while it is taken at the position 
of the primary star in the case of resolved double stars 
(for instance, Aitken 1967). 

At least five observations enable us to determine the parameters 
by using the least square method as 
\begin{eqnarray}
\left(
\begin{array}{ccccc}
\sum_j \bar x_j^4 & \sum_j \bar x_j^2 \bar y_j^2 
& 2\sum_j \bar x_j^3 \bar y_j & 2\sum_j \bar x_j^3 
& 2\sum_j \bar x_j^2 \bar y_j \\
\sum_j \bar x_j^2 \bar y_j^2 & \sum_j \bar y_j^4 
& 2\sum_j \bar x_j \bar y_j^3 & 2\sum_j \bar x_j \bar y_j^2 
& 2\sum_j \bar y_j^3 \\
\sum_j \bar x_j^3 \bar y_j & \sum_j \bar x_j \bar y_j^3 
& 2\sum_j \bar x_j^2 \bar y_j^2 & 2\sum_j \bar x_j^2 \bar y_j 
& 2\sum_j \bar x_j \bar y_j^2 \\
\sum_j \bar x_j^3 & \sum_j \bar x_j \bar y_j^2 
& 2\sum_j \bar x_j^2 \bar y_j & 2\sum_j \bar x_j^2 
& 2\sum_j \bar x_j \bar y_j \\
\sum_j \bar x_j^2 \bar y_j & \sum_j \bar y_j^3 
& 2\sum_j \bar x_j \bar y_j^2 & 2\sum_j \bar x_j \bar y_j 
& 2\sum_j \bar y_j^2 \\
\end{array}
\right)
\left(
\begin{array}{c}
\alpha \\
\beta \\
\gamma \\
\delta \\
\varepsilon \\
\end{array}
\right)
=
\left(
\begin{array}{c}
\sum_j \bar x_j^2 \\
\sum_j \bar y_j^2 \\
\sum_j \bar x_j \bar y_j \\
\sum_j \bar x_j \\
\sum_j \bar y_j \\
\end{array}
\right) , 
\label{ellipse2}
\end{eqnarray}
where the location of the star on the celestial sphere 
at the time of $t_j$ for $j=1, \cdots, n$ 
is denoted by $P_j=(\bar x_j, \bar y_j)$. 
The inverse of a matrix in the L. H. S. exists uniquely. 
In some case such as a large observational error, however, 
the determinant of the matrix can become nearly zero. 
Then, it would be more difficult to estimate the inverse matrix. 

The time interval between observations is denoted 
by $T(k, j)=t_k-t_j$ for $t_k > t_j$. 
We choose the Cartesian coordinates $(x, y)$
so that the observed ellipse given by Eq. ($\ref{ellipse1}$) 
can be reexpressed in the standard form as 
\begin{equation}
\frac{x^2}{a^2}+\frac{y^2}{b^2}=1 , 
\label{ellipse3}
\end{equation}
where $a\geq b$. 
The ellipticity $e$ is $\sqrt{1-b^2/a^2}$. 
There exists such a coordinate transformation as 
a combination of a translation and a rotation, 
which is expressed as 
\begin{eqnarray}
\left(
\begin{array}{c}
x \\
y \\
\end{array}
\right)
=
\left(
\begin{array}{cc}
\cos\Omega & \sin\Omega \\
-\sin\Omega & \cos\Omega \\
\end{array}
\right)
\left(
\begin{array}{c}
\bar x + \rho \\
\bar y + \sigma \\
\end{array}
\right) , 
\label{transformation}
\end{eqnarray}
where $\rho$ and $\sigma$ are given by 
\begin{eqnarray}
\left(
\begin{array}{c}
\rho \\
\sigma \\
\end{array}
\right)
=
\left(
\begin{array}{cc}
\alpha & \gamma \\
\gamma & \beta \\
\end{array}
\right)^{-1} 
\left(
\begin{array}{c}
\delta \\
\epsilon \\
\end{array}
\right) , 
\label{transformation2}
\end{eqnarray}
and $\Omega$ is determined by 
\begin{equation}
\tan 2\Omega=\frac{2\gamma}{\alpha-\beta} . 
\end{equation}
Then, $a$ and $b$ are given by 
\begin{eqnarray}
a=\sqrt{\frac{1+I}{A}} , 
\label{a}\\
b=\sqrt{\frac{1+I}{B}} ,
\label{b}
\end{eqnarray}
where $I\equiv \alpha\rho^2+\beta\sigma^2+2\gamma\rho\sigma$, 
$A\equiv \alpha\cos^2\Omega+\beta\sin^2\Omega
+\gamma\sin2\Omega$, and 
$B\equiv \alpha\sin^2\Omega+\beta\cos^2\Omega
-\gamma\sin2\Omega$. 

It has been recently shown (Asada et al. 2004) that the orbital
elements can be expressed explicitly as elementary functions of 
the locations of {\it four} observed points and their time intervals  
if astrometric measurements are done {\it without} any error. 
The key thing is that even after a Keplerian orbit is projected 
onto the celestial sphere, the law of constant-areal velocity 
(e.g., Goldstein 1980, Danby 1988, Roy 1988, Murray and Dermott 1999, 
Beutler 2004) still holds, where the area is swept 
by the line interval between the star and the projected common center 
of mass but not a focus of the observed ellipse. 
For example, let us consider four observed points 
$P_1$, $P_2$, $P_3$ and $P_4$ for $t_1<t_2<t_3<t_4$. 
The location $(x_e, y_e)$ of the projected common center 
is given (Asada et al. 2004) by
\begin{eqnarray}
x_e&=&-a \frac{F_1 G_2-G_1 F_2}{E_1 F_2-F_1 E_2} , 
\label{xe}\\
y_e&=&b \frac{G_1 E_2-E_1 G_2}{E_1 F_2-F_1 E_2} , 
\label{ye}
\end{eqnarray}
where 
\begin{eqnarray}
E_j&=&T(j+1, j)\sin u_{j+2}+T(j+2, j+1)\sin u_{j}
\nonumber\\
&&-T(j+2, j)\sin u_{j+1} , 
\label{E} \\
F_j&=&T(j+1, j)\cos u_{j+2}+T(j+2, j+1)\cos u_{j}
\nonumber\\
&&-T(j+2, j)\cos u_{j+1} , 
\label{F} \\
G_j&=&T(j+1, j)u_{j+2}+T(j+2, j+1)u_{j}-T(j+2, j)u_{j+1} , 
\label{G} 
\end{eqnarray}
and the eccentric anomaly in the observed ellipse is given 
by $u_j=\arctan(ay_j/bx_j)$. 
What Eqs. ($\ref{xe}$) and ($\ref{ye}$) tell us is 
the relative position of the projected common center of mass 
with respect to the center of the observed ellipse, 
that is the origin of our coordinates system.  
If one wishes to take the origin of the coordinates as 
another point such as the location of the primary star, 
$(x_e, y_e)$ must be shifted by the displacement from 
the center of the observed ellipse 
to the new point. 
The above solution is generalized to determination 
of an open orbit (Asada 2006).

\section{Extension to observational data} 
In reality, however, we have observational errors 
in each position measurement. 
We assume that these random errors obey the Gaussian distribution. 
When the errors vanish, any set of four points $P_p$, $P_q$, $P_r$ 
and $P_s$ $(t_p<t_q<t_r<t_s)$ among observational data must satisfy 
Eqs. ($\ref{xe}$) and ($\ref{ye}$), 
if one replaces $1\to p$, $2\to q$, $3\to r$ and $4\to s$. 
All of $a$, $b$, $E_j$, $F_j$ and $G_j$ in Eqs. ($\ref{xe}$) 
and ($\ref{ye}$) have been determined up to this point and 
$x_e$ and $y_e$ are the parameters to be estimated. 
We should note that Eqs. ($\ref{xe}$) and ($\ref{ye}$) 
are {\it linear} in parameters $x_e$ and $y_e$ to be 
determined by the least square method. 
Hence, $\chi^2$ becomes square in $x_e$ and $y_e$ 
so that the extremum of $\chi^2$ can tell us 
\begin{eqnarray}
x_e&=&-\frac{a}{{}_nC_4}\sum_j
\frac{F_j G_{j+1}-G_j F_{j+1}}{E_j F_{j+1}-F_j E_{j+1}} , 
\label{xen}\\
y_e&=&\frac{b}{{}_nC_4}\sum_j
\frac{G_j E_{j+1}-E_j G_{j+1}}{E_j F_{j+1}-F_j E_{j+1}} , 
\label{yen}
\end{eqnarray}
where the summation is taken for every set of four points 
$P_p$, $P_q$, $P_r$ and $P_s$ with a correspondence 
as $j\to p$, $j+1\to q$, $j+2\to r$ and $j+3\to s$ 
for defining $E_j$, $F_j$ and $G_j$ by Eqs. ($\ref{E}$)-($\ref{G}$), 
and the number of all the combinations is ${}_nC_4=n!/4!(n-4)!$.  
In practice, the number of summing, namely a denominator of 
Eqs. ($\ref{xen}$) and ($\ref{yen}$), can be reduced 
significantly from ${}_nC_4$ to $n/4$ in the case of $n=4 m$, 
if one considers only a set of four points 
as $(P_j, P_{j+m}, P_{j+2m}, P_{j+3m})$ 
for $j=1, 2, \cdots, m$. 
It is worthwhile to mention that each point $P_j$ appears 
only once in this case, and the reduction is useful 
when applied to a lot of data. 

The mapping between the observed ellipse and the Keplerian 
orbit parameterized by the major axis $a_K$ and 
the ellipticity $e_K$ is specified by the inclination angle $i$ 
and the ascending node $\omega$ (See Fig. 1). 

Given $a$, $b$, $x_e$ and $y_e$, one can analytically determine
the remaining parameters $e_K$, $i$, $a_K$ and $\omega$ 
in order as (Asada et al. 2004) 
\begin{eqnarray}
&&e_K=\sqrt{\frac{x_e^2}{a^2}+\frac{y_e^2}{b^2}} ,  
\label{eK}\\
&&\cos i=\frac12 (\xi - \sqrt{\xi^2-4}) , 
\label{cosi}\\
&&a_K=\sqrt{\frac{C^2+D^2}{1+\cos^2 i}} , 
\label{aK}\\
&&\cos 2\omega=\frac{C^2-D^2}{a_K^2 \sin^2 i} , 
\label{cos2omega}
\end{eqnarray}
where 
\begin{eqnarray}
&&C=\frac{1}{e_K}\sqrt{x_e^2 + y_e^2} , 
\label{C2}\\
&&D=\frac{1}{abe_K}\sqrt{\frac{a^4 y_e^2 + b^4 x_e^2}{1-e_K^2}} , 
\label{D2}\\
&&\xi=\frac{(C^2+D^2)\sqrt{1-e_K^2}}{ab} . 
\label{xi}
\end{eqnarray}

\section{Numerical test}
In this section, we verify the present formula by numerical computations 
for various values of the orbital parameters. 
In order to make a verification, we assume random errors 
in measurements of positions of the star and 
show that the orbital parameters are reconstructed: 
First, we assume a Keplerian orbit specified by $a_K$ and $e_K$ 
and a projection defined by $i$ and $\omega$. 
We prepare twenty seven sets of the parameters as 
$a_K=1$, $e_K=0.1, 0.3, 0.6$, $i=0, 30, 60$ deg. and 
$\omega=0, 30, 60$ deg. 

For each test, we pick up twelve points 
in the original Kepler orbit by assuming the same time interval 
between neighboring points for simplicity. 
Next, for assumed $i$ and $\omega$,  
we estimate the projected locations of the twelve points 
in the apparent ellipse plane. 
We add random errors to each point in order to imitate 
a ``measured position'' with observational errors (See Fig. 2). 
In order to produce such a simulated data, we assume that 
the absolute standard deviation of the random error 
in the position measurement is 0.001 
in the units of $a_K=1$.  
We apply our formula to the simulated data 
in order to determine the orbital elements, and 
make a comparison between the assumed value and the retrieved one. 
Table 1 shows a good agreement between them. 

For each parameter set, we perform a hundred of numerical runs 
for a statistical treatment. 
We define an error as the absolute standard deviation as 
$\Delta a=\sqrt{<(a_K'-a_K)^2>}$,  $\Delta e=\sqrt{<(e_K'-e_K)^2>}$, 
$\Delta i=\sqrt{<(i'-i)^2>}$ and 
$\Delta \omega=\sqrt{<(\omega'-\omega)^2>}$, 
where the prime denotes values retrieved 
by using the present formula and the square bracket 
denotes a mean over each set. 
In Table 1, asterisk ($\ast$) indicates that the formula 
does not always give a real value but a complex one. 
The case of complex numbers occurs because a large error 
in measurements of the positions in the apparent ellipse plane 
causes an anomaly in the apparent motion such as 
an apparent clockwise motion even if a true one is anti-clockwise. 


It is worthwhile to make a comment on Table 1:  
In the case of no inclination as $i=0$, 
the ascending node does not exist, so that 
an error in retrieving $\omega$ can be apparently quite large, 
though it is harmless. 

\section{Conclusion}
In this paper, as a simplified method to determine a binary orbit, 
we extend the solution considering observational data. 
We expect that the present formula will be used 
for orbit determinations by some planned astrometric space 
missions such as 
SIM\footnote{http://sim.jpl.nasa.gov/} (Shao 2004), 
GAIA\footnote{http://astro.estec.esa.nl/GAIA/} 
(Mignard 2004, Perryman 2004) and 
JASMINE\footnote{http://www.jasmine-galaxy.org/} 
(Gouda et al. 2004), 
which will observe with the accuracy of possibly a few 
micro arcseconds a number of binaries 
whose components are a visible star and a companion 
ranging from a black hole to an extra-solar planet. 
Our numerical computations correspond to a case that 
by these missions we observe a binary system 
at a distance of 1 kpc from us, 
where the semimajor axis of the primary's orbit is around 1AU.  



\acknowledgments

We would like to thank Professor N. Gouda for useful information 
on future astrometry missions. 
We would like to thank Professor M. Kasai and Professor S. Kuramata 
for continuous encouragement.

\clearpage



\begin{figure}
\plotone{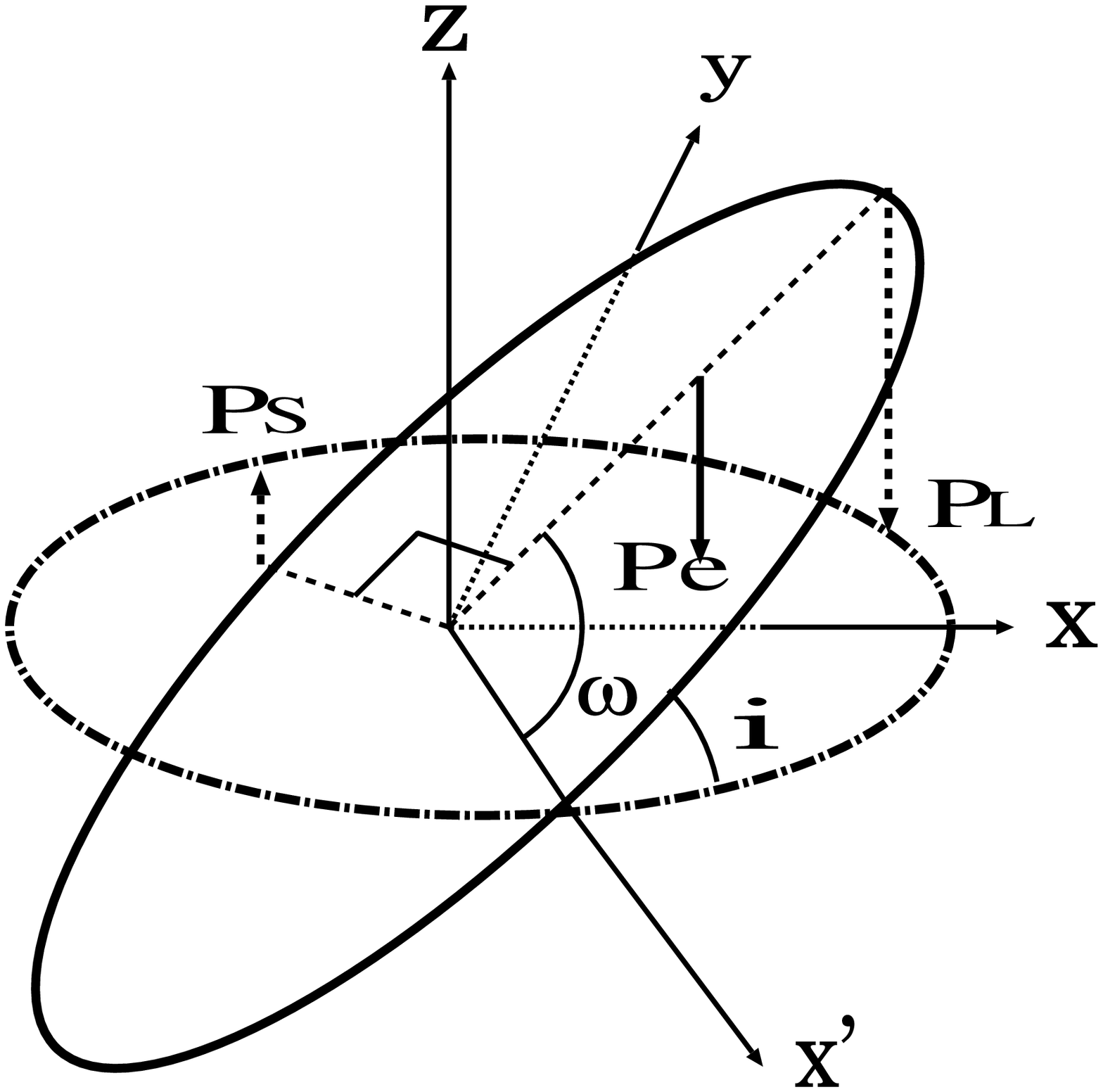}
  \caption{A relation between the Keplerian ellipse 
and the observed one. The thick solid curve shows the Keplerian 
ellipse and the dash-dotted curve shows the observed one 
projected onto a celestial sphere denoted by $x$-$y$ plane. 
The line of sight is along the $z$-axis.
The common center of mass is projected into the point 
denoted by $P_e$ in the apparent ellipse plane. 
The locations of the intersection of the observed ellipse 
and the projected major and minor axes are denoted by 
$P_L$ and $P_S$, respectively.  
}
\end{figure}

\clearpage


\begin{figure}
\plotone{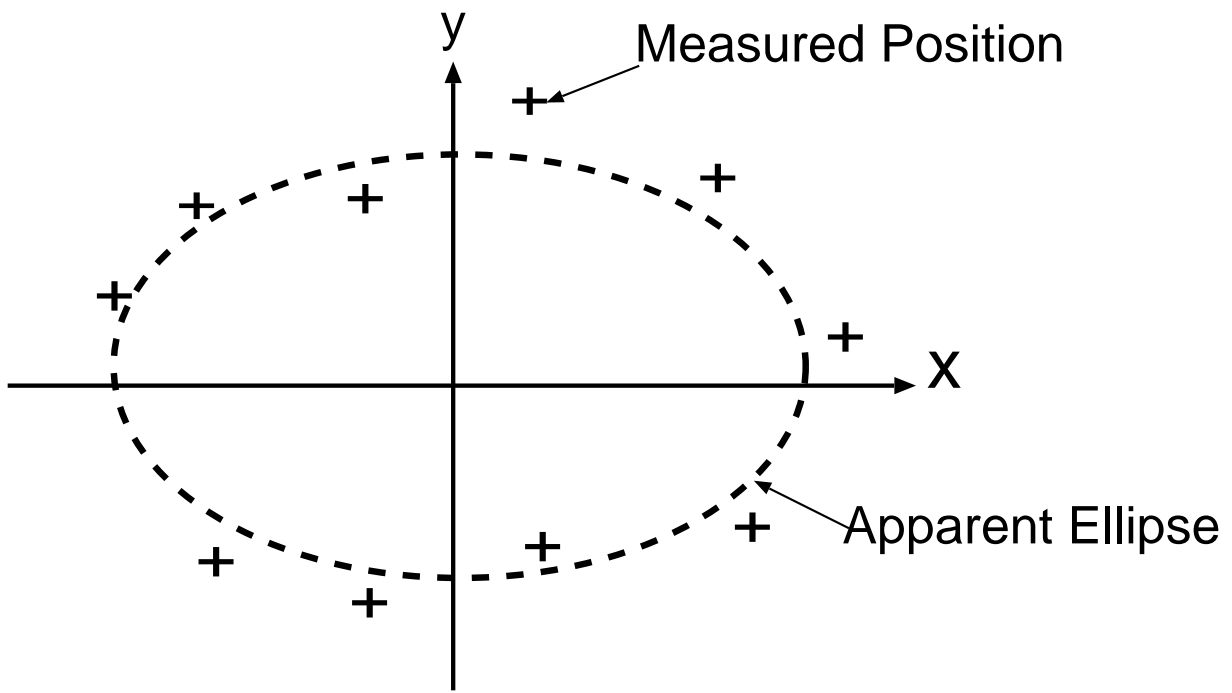}
  \caption{A schematic figure of a simulation. 
The sign ``+'' denotes a measured position in the simulations. 
The dashed curve is an apparent ellipse. 
}
\end{figure}

\clearpage

\begin{deluxetable}{crrrr}
\tabletypesize{\scriptsize}
\tablecaption{Reconstructing the parameters for a simulated data} 
\tablewidth{0pt}
\tablehead{
\colhead{$a_k-e_k-i-\omega$} & \colhead{$\Delta{a}$} & \colhead{$\Delta{e}$} 
& \colhead{$\Delta{i}$ [deg.]} & \colhead{$\Delta{\omega}$ [deg.]} 
}
\startdata
1-0.1- 0- 0 & 0.00113 & 0.00271 & 3.27 & 50.9\\
1-0.1- 0-30 & 0.00106 & 0.00258 & 3.16 & 24.7\\
1-0.1- 0-60 & * & * & * & * \\
1-0.1-30- 0 & 0.000866 & 0.00305 & 0.116 & 1.34\\
1-0.1-30-30 & 0.00110 & 0.00296 & 0.154 & 1.30\\
1-0.1-30-60 & 0.00110 & 0.00343 & 0.122 & 1.04\\
1-0.1-60- 0 & 0.00112 & 0.00450 & 0.0574 & 2.08\\
1-0.1-60-30 & 0.00193 & 0.00544 & 0.0951 & 2.20\\
1-0.1-60-60 & 0.00141 & 0.00501 & 0.0605 & 1.90\\
1-0.3- 0- 0 & 0.00180 & 0.00497 & 4.29 & 47.5\\
1-0.3- 0-30 & 0.00166 & 0.00516 & 4.29 & 27.8\\
1-0.3- 0-60 & 0.00193 & 0.00555 & 4.52 & 28.4\\
1-0.3-30- 0 & 0.000933 & 0.00518 & 0.224 & 0.943\\
1-0.3-30-30 & 0.00175 & 0.00542 & 0.317 & 0.719\\
1-0.3-30-60 & 0.00142 & 0.00597 & 0.164 & 0.449\\
1-0.3-60- 0 & 0.00157 & 0.00884 & 0.122 & 1.17\\
1-0.3-60-30 & 0.00238 & 0.00856 & 0.150 & 0.832\\
1-0.3-60-60 & 0.00227 & 0.00797 & 0.0888 & 0.715\\
1-0.6- 0- 0 & 0.0105 & 0.0137 & 9.16 & 54.6\\
1-0.6- 0-30 & 0.00977 & 0.0147 & 9.24 & 34.8\\
1-0.6- 0-60 & 0.0131 & 0.0150 & 9.48 & 33.3\\
1-0.6-30- 0 & 0.00240 & 0.0168 & 1.67 & 2.37\\
1-0.6-30-30 & 0.00374 & 0.0172 & 1.32 & 2.48\\
1-0.6-30-60 & 0.00953 & 0.0150 & 0.623 & 2.54\\
1-0.6-60- 0 & 0.00400 & 0.0279 & 0.919 & 1.68\\
1-0.6-60-30 & 0.00614 & 0.0287 & 0.765 & 0.966\\
1-0.6-60-60 & 0.0117 & 0.0191 & 0.256 & 0.586\\
\enddata


\end{deluxetable}






\end{document}